\documentclass[12pt,letterpaper]{article}
\usepackage{graphicx}
\usepackage{amsmath}

\newcommand{\ds}{\displaystyle}
\DeclareMathOperator{\erfc}{erfc}

\begin{document}
   \title{Tricriticality and Reentrance in a \textit{Naive} Spin-Glass Model}
   \author{Francisco A. da Costa \\
           Departamento de F\'{\i}sica Te\'orica e Experimental \\
           Universidade Federal do Rio Grande do Norte \\
           Caixa Postal 1641 \\
           59072-970 Natal, RN, Brazil \\
           E-mail: fcosta@dfte.ufrn.br}

   \date{\today}

\maketitle

   \begin{abstract}
     In this paper a spin-1 spin-glass model under the presence of a uniform crystal field is investigated. It is shown that the model presents both continuous and first-order phase transition separated by a tricritical point. The phase diagram is obtained within the replica-symmetric solution and exhibits reentrance phenomena at low temperatures. Possibly it is the simplest model which can describe inverse freezing phenomena.
   \end{abstract}

\vspace{.75cm}

Tricritical behavior and reentrance phenomena associated with first-order transitions in disordered systems have been the subject of recent studies. For instance, investigations of inverse freezing phenomena, where the ordered phase is more entropic than the disordered one, have been conducted by several authors \cite{plc10,leu07,sel06,cri05,sch05,sch04}. Such nonusual behavior has also been observed in fermionic disordered systems (see \cite{mag10} and references therein). More recently, a spin model for strain glass has been used to account for the effects of disorder in ferroelastics which can display reentrance phenomena \cite{vas10}. The models involved in those investigations are closely related to the Ghatak-Sherrington (GS) model \cite{gha77}, which has been intensively studied since its introduction \cite{cri05, lag82, mot85, cys94, cri02, cri04}.

The Ghatak-Sherrington model \cite{gha77} is a generalization of the widely known infinite-range Sherrington-Kirkpatrick (SK) model for a spin glass (SG) \cite{sk75} with arbitrary spin $S > 1/2$ and inclusion of a uniform crystal field. For integer spin $S$ the GS model displays both first-order and continuous transitions. The crystal field ($D$)-temperature ($T$) phase diagram for $S=1$ has a continous transition line which meets a first-order transitions (FOT) line at a tricritical point. These lines separates the paramagnetic from the spin-glass phase. One of the most interesting feature of the $S=1$ SG model is the appearance of reentrance effects occuring at low temperatures which has been recently associated with inverse freezing phenomena \cite{plc10,leu07,sel06,cri05,sch05,sch04}. 

Although it is widely accepted that the correct mean-field solution for infinite-range spin-glass models is given by the Parisi \textit{Ansatz} \cite{par1,par2,par3}, the replica-symmetric (RS) solution gives an initial clue to general topologies of phase diagrams. For the GS model the RS solution has revelead a number of difficulties in the low temperature region where first-order transtions take place \cite{cys94}. Inside this region we can find up to three distincts paramagnetic solutions. However, only one of these solutions is stable with respect to replica-symmetry breaking (RSB) fluctuations. On the other hand, we may find up to four distincts spin-glass solutions in the same region which, unlike the paramagnetic case, are all unstable with respect to RSB. Therefore, even within the RS solution treatment we are faced with the problem of choosing the most adequate SG solution in order to determine the FOT line. We believe that this fact is at the origin of some controversies with respect to the location of the FOT line predicted by the RS solution \cite{gha77, lag82, mot85}. Recently we have seen some progress with respect to the correct location of the FOT line for the GS model \cite{cri02,cri04}. There remain, however, some points which need to be clarified about the low temperature behavior of the GS model. 

Another route of investigation on the mean-field behavior of spin-glasses  is the Thouless-Anderson-Palmer (TAP) approach which avoids the replica method \cite{tap}. From the TAP equations one can obtain a simpler set of equations by excluding the so-called Onsager reaction-field term so that one gets the \textit{naive} mean-field equations. Bray, Sompolinsky and Yu (BSY) \cite{bsy86} introduced an exactly soluble infinite-range SG model which is also a generalization of the SK model. For the BSY model one can obtain exactly the \textit{naive} mean-field equations. These authors also studied their model by means of the replica method and noted some interesting features. For instance, the RS solution describes a SG phase whose entropy is always non-negative for finite temperatures and vanishes at zero temperature, in contrast to what is observed in the RS solution to the SK model. In spite of this, the RS solution to the BSY model is unstable in the whole SG region. Due to lack of the Onsager term the BSY predicts a critical temperature that is twice the result obtained for the SK model. Notwithstanding this, the low temperature behavior of both BSY and SK models are qualitatively the same. In particular they have the same zero-tempareture properties.

In order to gain further understanding about tricritical behavior and reentrant effects in spin-glass systems we consider a BSY version of the GS model. Our analysis is based on the replica approach, since we want to compare our findings with the known results obtained for the GS model in previous studies.

The model consists of a set of $m$ classical spin-1 variables $S_{ia} ~(a = 1, \dots , m)$ located at each site $i = 1, \dots , N$. The Hamiltonian is given by
   \begin{equation} 
     \label{ham1}
     \mathcal{H} = - \frac{1}{2 m} \sum_{(i,j)} \sum_{a,b=1}^{m} J_{ij}  S_{ia} S_{jb} + D \sum_{i=1}^{N}\sum_{a=1}^{m} S_{ia}^2 ,
   \end{equation}

\noindent   
where $S_{ia} = \pm 1, 0$; the $(i,j)$ sum is over all
distinct pairs of sites; the exchange interactions $J_{ij}$  are quenched random variables with the Gaussian distribution 

   \begin{equation} \label{gd}
   P(J_{ij}) = \left(\frac{N}{2 \pi J^2}\right)^{1/2} \exp \left( - \frac{N J_{ij}^2}{2J^2}  \right),
   \end{equation}

\noindent
and $D$ represents the effect of a uniform crystal field anisotropy term. Several known cases are recovered in special limits: (i) $m=1$  recovers the Ghatak-Sherrington model; (ii) for $D \to -\infty$ the BSY model is re-obtained, including the SK model for $m=1$. 

Since we are interested in the $m \to \infty$ limit, the quenched free energy  per spin is given by

\begin{equation} \label{fed1}
-\beta f = \lim_{N \to \infty} \lim_{m \to \infty} \frac{1}{N m} \left\langle \ln Z \right\rangle_J ,
\end{equation}

\noindent
where $\left\langle \right\rangle_J$ denotes the configuration average over the disorder. In order to proceed further the use of the replica method is introduced through the identity $\ln Z = \lim_{n \to 0} (Z^n - 1)/n $. Therefore, the averaged free energy density may be expressed as 

\begin{equation} \label{fed2}
\beta f = \lim_{n \to 0} \lim_{m \to \infty} \frac{1}{m n} \min \phi_{mn} (\{q_{\alpha \beta}\}) ,
\end{equation}

\noindent
where

\begin{equation}
\phi_{mn} (\{q_{\alpha \beta}\}) = \frac{1}{4} m^2 (\beta J)^2 
\sum_{\alpha, \beta = 1}^{n} q_{\alpha \beta}^2 - \ln \normalfont{Tr}_{\{S^{\alpha}_{a}\}} \exp ( \mathcal{H}_{mn} ) ,
\end{equation}
\noindent 
and

\begin{equation} \label{rham1}
\mathcal{H}_{mn} = \frac{1}{2} (\beta J)^2 \sum_{\alpha, \beta = 1}^n \sum_{a,b = 1}^{m} q_{\alpha \beta} S^{\alpha}_{a} S^{\beta}_{b} - \beta D \sum_{\alpha, a} (S^{\alpha}_{a})^2 .
\end{equation}

The condition for $\phi_{mn} (\{q_{\alpha \beta}\})$ to be an extremum with respect to $q_{\alpha \beta}$ yields

\begin{equation}
q_{\alpha \beta} = \frac{1}{m^2} \sum_{a,b=1}^{m} 
\langle S^{\alpha}_{a} S^{\beta}_{b} \rangle_{mn}
\end{equation}

\noindent
where $\langle S^{\alpha}_{a} S^{\beta}_{b} \rangle_{mn}$ indicates the thermal average with respect to the replica hamiltonian (\ref{rham1}).

The replica-symmetric solution can be considered now. As in the BSY model there is a non-trivial diagonal $q_{\alpha \alpha}$ and the requirement of a finite susceptibility leads us to consider the RS \emph{Ansatz} in the form:

\begin{equation} \label{rs1}
q_{\alpha \beta} = q \quad (\alpha \neq \beta), \qquad q_{\alpha \alpha} = q + \bar{\chi}/m .
\end{equation}

Substituting (\ref{rs1}) into (\ref{fed2}) the free energy per spin is obtained

\begin{equation} \label{rsf}
\beta f = \frac{1}{2} (\beta J)^2 \bar{\chi} q - \left \langle -\frac{1}{2}   (\beta J)^2 \bar{\chi} m^2(x) + \ln z(x) \right \rangle_x , 
\end{equation}

\noindent
where 

\begin{equation} \label{zx}
z(x) = 1 + 2e^{-\beta D} \cosh(\beta J \sqrt{q} x + \beta^2 J^2 \bar{\chi} m(x)) ,
\end{equation}

\noindent
and

\begin{equation} \label{mx}
m(x) = \displaystyle{\frac{2 \sinh(\beta J \sqrt{q} x + \beta^2 J^2 \bar{\chi} m(x))}{e^{\beta D} + 2 \cosh(\beta J \sqrt{q} x + \beta^2 J^2 \bar{\chi} m(x))}} .
\end{equation}

The notation has been simplified by introducing

$$\langle \mathcal{O}(x) \rangle_x  = \int_{-\infty}^{\infty} \mathcal{O}(x) e^{-x^2/2} \frac{dx}{\sqrt{2 \pi}}. $$

The equilibrium equations $\partial f/\partial q = 0 = \partial f/\partial \bar{\chi}$ yield

\begin{equation} \label{ee1}
q = \langle m^2(x) \rangle_x , 
\end{equation}

\begin{equation} \label{ee2}
\beta \bar{\chi} =  \frac{1}{\sqrt{q}} \langle x m(x) \rangle_x .
\end{equation}

An integration by parts allows to re-write (\ref{ee2}) as

\begin{equation} \label{ee3}
\bar{\chi} =  \left\langle \frac{ p(x) - m^2(x)}{1 - (\beta J)^2 \bar{\chi} (p(x) - m^2(x))} \right\rangle_x ,
\end{equation}

\noindent
where

\begin{equation} \label{px}
p(x) = \displaystyle{\frac{2 \cosh(\beta J \sqrt{q} x + \beta^2 J^2 \bar{\chi} m(x))}{e^{\beta D} + 2 \cosh(\beta J \sqrt{q} x + \beta^2 J^2 \bar{\chi} m(x))}} .
\end{equation}

The set of Eqs. (\ref{rsf}), (\ref{ee1}) and (\ref{ee2}) determine the phase diagram which presents paramagnetic and spin-glass phases. It should be mentioned that for numerical purpose Eq. (\ref{ee3}) is more appropriate than Eq. (\ref{ee2}).

The paramagnetic phase is described by $ q= 0 $ and 

\begin{equation} \label{ee4}
\bar{\chi} =  \frac{1 - \sqrt{1- 4 \beta^2 J^2 p^2}}{2 \beta^2 J^2 p} , \qquad p = \frac{2}{e^{\beta D} + 2}  .
\end{equation}

The simple form of $p$ in the last equation shows that it is a single valued function both in terms of inverse temperature $\beta$ and anisotropy crystal field $D$. It is important to recall that in the Ghatak-Sherrington model the behavior of the corresponding term is more subtle \cite{cys94}. As a matter of fact in the GS case $p$ represents a true order parameter and can display up to three distinct solutions. Thus we have to find an additional criterion in order to determine the thermodynamically stable paramagnetic solution. In the present case $p$ can be regarded as a mere density and a simple analysis shows that the paramagnetic solution is physically acceptable as long as the condition

\begin{equation} \label{ee5}
 \frac{2 \beta^2 J^2}{e^{\beta D} + 2} < 1
\end{equation}

\noindent
holds.

Let us set $J = 1$ and consider the $D - T$ phase diagram, where $T = 1/\beta$. From Eq. (\ref{ee5}) the paramagnetic is stable at high temperatures and is bordered by the line

\begin{equation} \label{pmb}
 D = T \ln \left[ \frac{2(2-T)}{T}\right] .
\end{equation}

At low temperatures there is a spin-glass phase with $q > 0$. Expanding Eq. (\ref{ee1}) for small $q$ one finds

\begin{equation} \label{exp1}
	q = a q + b q^2 + O(q^3) 
\end{equation}

\noindent
where

\begin{equation}
	a = \frac{p T}{T^2 - \bar{\chi} p} ,
\end{equation}

\noindent
and

\begin{equation}
	b = \frac{1}{9} \frac{(1 - 3p)}{p^3} T a^4 .
\end{equation}

Thus the spin-glass phase exists as long as $a < 1$ and $b > 0$. From the above expansion one finds a tricritical point given by $a = 1$ and $b = 0$. Therefore the phase diagram consists of a continuous transition line given by Eq. (\ref{pmb}) as long as $ T > 2/3 $. For $ 0 \le T < 2/3 $ there is a region of coexisting paramagnetic and spin-glass solutions and one has a first-order transition which can be numerically determined by equating the corresponding free-energy densities of these solutions. The continuous and first-order transition lines meet at the tricritical point given by

\begin{equation}
	T = 2/3, \quad D = 4 \ln 2 / 3 = .924169 \dots ,
\end{equation}

\noindent 
which should be compared with the correponding results for the tricritical point found for the GS model: $T = 1/3, ~D = 1/2 + 2(\ln 2)/3 = 0.962098 \dots$ \cite{gha77}.

At zero temperature the first-order transition can also be easily found. First one notes that $\lim_{T \to 0} \beta \bar{\chi} = \chi $ is finite. From Eq. (\ref{ee1}) one finds

\begin{equation} \label{q0}
	q = 2 \int_{x^{*}}^{\infty} \ds{e^{-x^2/2}} \frac{dx}{\sqrt{2 \pi}} = \erfc (x^{*}/\sqrt{2}) ,
\end{equation}

\noindent
where 

\begin{equation}
	x^{*} = \frac{2D - \chi}{2\sqrt{q}} ,
\end{equation}

\noindent
and $\erfc$ is the usual complementary error function. The above expression is valid for $x^{*} > 0$. In this regime one also finds

\begin{equation} \label{chi0}
	\chi = \left(\frac{2}{\pi q}\right)^{1/2} e^{-{x^{*}}^{2}/2} ,
\end{equation}

\noindent
and, for the spin-glass free-energy density $f_0$,

\begin{equation} \label{fsg0}
	f_{0} = D q - \left(\frac{2 q}{\pi}\right)^{1/2} e^{-{x^{*}}^{2}/2}.
\end{equation}

For negative values of $x^{*}$ we obtain, at the absolute zero of temperature, $q = 1$, $\chi = \left(2/\pi\right)^{1/2}$ and $ f = - \left(\pi/2\right)^{1/2} $. Since from Eq. (\ref{rsf}) the free energy of the paramagnetic solution at $T = 0$ is zero for $D > 0$, the first-order transitions at $T=0$ is found by imposing, $f_0 = 0$ from which follows, after some simplifications with the help of Eqs. (\ref{q0}) and (\ref{chi0})

\begin{equation} \label{fol0}
 x^{*} \erfc (x^{*}/\sqrt{2}) = \frac{1}{\sqrt{2 \pi}} e^{-{x^{*}}^{2}/2} .
\end{equation}

The above equation is exactly the same obtained previously for the replica symmetric solution \cite{cys94} to the GS model. The numerical solution to this equation is $x^{*} = 0.612003 \dots$, from which results, jointly with Eqs. (\ref{q0}) and (\ref{chi0}), $q = 0.540535 \dots$ and $\chi = 0.899003 \dots $, respectively. These numerical results allow us to determine the location of the first-order transition at $ T = 0$:

\begin{equation}
	 D_0 = D (T = 0) = 0.899033 \dots ~.
\end{equation}

Since the present model as well as the GS model have the same ground state this is an expected result of general validity. Thus, as in the GS model a stable RSB solution should give a slightly lower value for $D_0$, and so the reentrance effect must be enhanced. In fact, a previous numerical study of the \textit{naive} mean-field equations for $T=0$ showed that $D_0 \approx 0.86 $ \cite{cos00}.

For $0 < T < 2/3$ the first-order transition line can be obtained by numerically solving the equilibrium equation and requiring that the spin-glass and paramagnetic solution have the same free energy. The resulting phase diagram is depicted in Fig. 1. As in the GS model case there is a reentrance to the paramagnetic phase at low temperatures. However, one notices that as $ T \to 0 $ there is no new transition to the SG phase in the vicinity of $D_0$ as was found in the replica solution to the Ghatak-Sherrington model \cite{cys94}. This is a direct consequence of the vanishing of the spin-glass entropy as can be verified by an analysis of the Clausius-Clayperon equation along the FOT line.

In spite of the vanishing of the spin-glass entropy at $T=0$ for any value of $D$, a stability analysis along the lines pioneered by de Almeida and Thouless \cite{at78} shows that the replica-symmetric spin-glass solution is always unstable. Again, the numerical analysis of the stability conditions becomes  easier than in the corresponding Ghatak-Sherrington case since we did not find any evidence of complex eigenvalues for the replica stability matrix. The instability is signaled by the non-positiveness of the replicon eigenvalue

\begin{equation}
	 \lambda_R = T^2 - \langle (p(x) - m^2(x))^2 \rangle_x .
\end{equation}

Therefore, a correct description of the spin-glass phase requires a complete solution to the corresponding Parisi's equations as have been done for the GS model \cite{cri02,cri04}. We can anticipate that in such replica-symmetry breaking treatment the main features of the phase diagram would not be modified. The continuous transition line as well as the location of the tricritical point will not be changed, but a slight modification in the location of the FOT line is expected in a similar way to what happens in the GS model, increasing teh reentrance effect for $T \approx 0$.

In conclusion, a naive version of the Ghatak-Sherrington model was investigated by the replica approach. The phase diagram was determined within the replica-symmetric \emph{Ansatz}. Both analytical and numerical results shows that the replica-symmetric solution to the present model is simpler to analyse than the corresponding Ghatak-Sherrington model. In spite of this, the phase diagram for both models share several common features. At high temperatures there is continuous transition line from the paramagnetic to the spin-glass phase, while at low temperatures there is a line of first-order transitions. These two lines meet at a tricritical point. The present model also exhibits a reentrance from the spin-glass to the paramagnetic phase. We believe that the present model could be useful to further investigations on inverse freezing phenomena.

The author thanks to Prof. S\'{\i}lvio Salinas for useful comments.

\newpage
\setlength{\unitlength}{1cm}
\begin{picture}(-100,-100)
\put(3,-8.0){\framebox(7.5,9)}
\linethickness{0.3mm}
\multiput(4.875,-8)(1.875,0){3}{\line(0,1){.1}}
\linethickness{0.3mm}
\multiput(3,-6.5)(0,1.5){5}{\line(1,0){.1}}
{\thicklines
\qbezier(8.0,-8.0)(7.875,-7.25)(7.6875,-6.275)
\qbezier(7.6875,-6.275)(7.5,-5.45)(7.3125,-4.625)
\qbezier(7.3125,-4.625)(7.125,-3.950)(6.9375,-3.35)
\qbezier(6.9375,-3.35)(6.75,-2.825)(6.5625,-2.375)
\qbezier(6.5625,-2.375)(6.375,-2.)(6.1875,-1.655)
\qbezier(6.1875,-1.655)(6.,-1.415)(5.8125,-1.2275)
\qbezier(5.8125,-1.2275)(5.71875,-1.16)(5.625,-1.111)
\qbezier(5.625,-1.111)(5.55,-1.08274)(5.50125,-1.0685)}
\put(5.50125,-1.0685){\circle*{.15}}
\put(3,-1.25){\circle*{.075}} 
\put(3.1875,-1.25){\circle*{.075}} 
\put(3.375,-1.25){\circle*{.075}} 
\put(3.5625,-1.25){\circle*{.075}} 
\put(3.75,-1.23){\circle*{.075}} 
\put(3.9375,-1.2125){\circle*{.075}}  
\put(4.125,-1.195){\circle*{.075}}  
\put(4.3125,-1.175){\circle*{.075}}  
\put(4.5,-1.15){\circle*{.075}}  
\put(4.6875,-1.105){\circle*{.075}}  
\put(4.875,-1.08){\circle*{.075}}  
\put(5.0625,-1.05){\circle*{.075}}  
\put(5.25,-1.035){\circle*{.075}}  
\put(5.4375,-1.055){\circle*{.075}}  
\put(6.75,-9.00){$T$}
\put(2.88,-8.5){0}
\put(4.625,-8.5){0.5}
\put(6.5,-8.5){1.0}
\put(8.35,-8.5){1.5}
\put(10.25,-8.5){2.0} 
\put(1.75,-3.5){$D$}
\put(2.5,-8.15){0}
\put(2.25,-5.15){0.4}
\put(2.25,-2.15){0.8} 
\put(2.25,0.85){1.2}
\put(5.5,-.95){\small T}
\put(3.07,-1.7){\small B}
\put(8.10,-7.9){\small A}
\put(4.6,-5.0){\small SG}
\put(8.1,-1.0){\small PM}
%
%
\end{picture}
\vskip 9.5cm
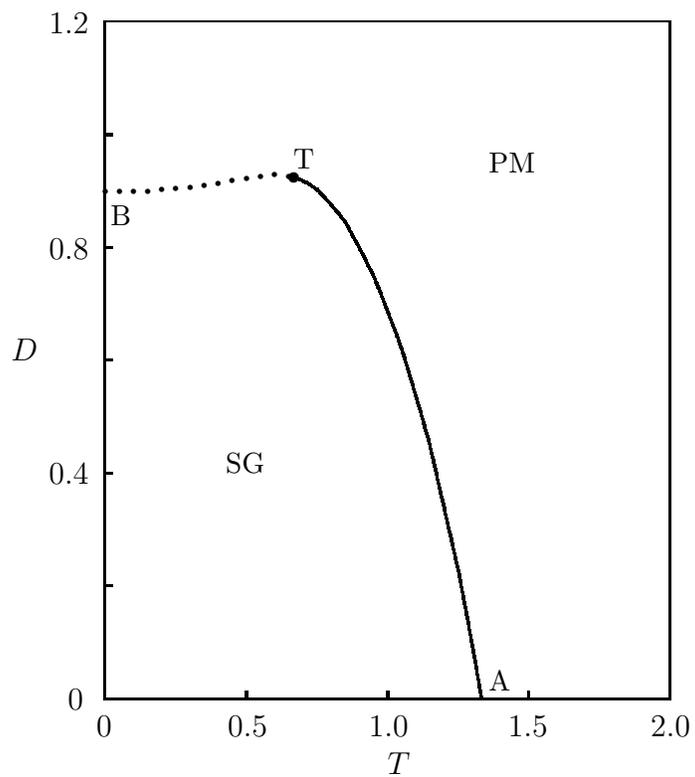
{\small \textbf{Figure 1} Phase diagram obtained within the 
replica-symmetric approximation. The full AT 
 curve is the line of continuous transitions and the
broken curve BT is the line of first-order transitions.
These two curves meet at the tricritical point T.}

\newpage

\end{document}